# Trichromatic and Tri-polarization-channel Holography with Non-interleaved Dielectric Metasurface

Yueqiang Hu[1], Ling Li[1], Min Meng[2,3], Lei Jin[4], Xuhao Luo[1], Yiqin Chen[1], Li, Xin[1], Hanbin Wang[2,3], Yi Luo[2], Cheng-Wei Qiu[4*], and Huigao Duan[1*]

[1]State Key Laboratory of Advanced Design and Manufacturing for Vehicle Body, College of Mechanical and Vehicle Engineering, Hunan University, Changsha 410082, P.R. China

[2]Microsystem & Terahertz Research Center, China Academy of Engineering Physics (CAEP), Chengdu 610200, P.R. China

[3]Institute of Electronic Engineering, China Academy of Engineering Physics (CAEP), Mianyang 621900, P.R. China

[4]Department of Electrical and Computer Engineering, National University of Singapore, 4 Engineering Drive 3, Singapore 117583, Republic of Singapore

*Corresponding authors. Email: duanhg@hnu.edu.cn and chengwei.qiu@nus.edu.sg

**Abstract:**

Metasurfaces hold great potentials for advanced holographic display with extraordinary information capacity and pixel sizes in an ultrathin flat profile. Dual-polarization channel to encode two independent phase profiles or spatially multiplexed meta-holography by interleaved metasurfaces are captivated popular solutions to projecting multiplexed and vectorial images. However, the intrinsic limit of orthogonal polarization-channels, their crosstalk due to coupling between meta-atoms, and interleaving-induced degradation of efficiency and reconstructed image quality set great barriers for sophisticated meta-holography from being widely adopted. Here we report a non-interleaved TiO$_2$ metasurface holography, and three distinct phase profiles are encoded into three orthogonal polarization bases with almost zero crosstalk. The corresponding three independently constructed intensity profiles are therefore assigned to trichromatic (RGB) beams, resulting in high-quality and high-efficiency vectorial meta-holography in the whole visible regime. Our strategy presents an unconventionally advanced holographic scheme by synergizing trichromatic colors and tri-polarization channels, simply realized with a minimalist non-interleaved metasurface. Our work unlocks the metasurface's potentials on massive information storage, polarization optics, polarimetric imaging, holographic data encryption, etc.

**Introduction**

Subwavelength metal or dielectric meta-units arranging in a two-dimensional plane form a metasurface, which can almost arbitrarily manipulate the wavefront such as amplitude, phase and polarization[1-3]. The complementary metal oxide semiconductor (CMOS) compatible fabrication process enables various metasurfaces components in the optical range that are more compact and perform better than conventional refractive optics, such as abnormal reflector/refractor[1, 4-6], color filter[7, 8], metalens[3, 9-11], meta-hologram[12-14], quantum metasurfaces[15-17] and multifunctional metasurfaces[18-21]. Multichannel metasurface-based devices have been extensively studied. One important application is to reconstruct full-color images. The crosstalk among different wavelengths is taken care of mainly by two approaches: optimized phase retrieval algorithm and judiciously optimized meta-atoms. The first approach enables a single phase profile to generate a color image by introducing position information in the phase retrieval processes to suppress the crosstalk among different operational wavelengths in the observation region[22-24]. However, unwanted images will be induced in other regions. On the other hand, the second approach is to provide wavelength-dependent phase profiles via carefully tailored meta-atoms[14, 25-29], which can support wavelength-dependent responses. The meta-atom subarrays specifically designed for each given wavelength can be interleaved with the others to provide independent phase profiles for different working wavelengths, respectively, which in turn deteriorates the efficiency and reconstructed image quality due to inevitably enlarged periodicity necessary for interleaving. Moreover, the meta-atom array could be carefully tailored so as to induce additional wavelength-dependent phase variations, on top of the intrinsic single-phase profile[27]. This mechanism could allow full-color holographic images, but material's limited dispersion impairs the independence of the phase profiles at individual wavelengths.

Here, we experimentally demonstrate a novel tri-polarization-channel and trichromatic holography by a non-interleaved $TiO_2$ metasurface, which imparts three distinct phase profiles at red, green and blue lights, respectively. Such three phase profiles are encrypted into one rectangular meta-atom and vectorial full-color meta-holography can be independently reconstructed through different combinations of input-output polarizations (H/V, 45°, and RCP/LCP). Trichromatic components are coupled into the three independent polarization channels to achieve full-color vectorial holography. The crosstalk between channels is thus negligible in principle, and there are no extra images in other spatial locations. Our approach establishes a new paradigm to achieve high-efficiency full-color vectorial meta-holography with nearly no crosstalk, at the price of minimalist and non-interleaved metasurface design.

**Results**

**Design and fabrication of the metasurfaces**

Figure 1 shows the schematic of the metasurface to achieve vectorial full-color holography. We first explore different numbers of channels of polarization-dependent holographic metasurfaces by considering different flexibility of the Jones matrix as shown in Fig.1a, b. The metasurfaces are realized by array of TiO$_2$ rectangular nanopillars with three independent structure parameters: length ($D_1$), width ($D_2$) and in-plane orientation angle ($\theta$). By controlling the length and width with fixed orientation angle, two independent phases can be encoded into a single nanopillar with two orthogonal incident polarization states. Therefore, two independent far field holographic images ('HUNAN' and 'UNIVERSITY' shown in Fig. 1a) can be switched with different polarization inputs. Via fully utilizing the parametric degrees of freedom by further considering the orientation angle, three kinds of independent information can also be realized with different combinations of input and output polarization states. As a result, the metasurfaces can reconstruct up to seven polarization-dependent information combinations (i.e., combinations of three independent 'X', 'Y' and 'Z' as shown in Fig. 1b). In addition, it is worth noting that the proposed metasurfaces are broadband in the visible, which enables broadband or multicolor applications. Due to the chromatic aberration, the size and field of view (FOV) of holographic images of different wavelengths projected on the same plane are proportional to the wavelengths. Using the broadband nature, we can ingeniously couple the three primary color wavelengths of grayscale image information into three independent polarization channels and then match the information of three channels by a pre-compensation algorithm to reconstruct a full-color holographic grayscale image, as shown by Fig.1c. Since the information is coupled into three independent input/output polarization combinations, the crosstalk images of the other channels are naturally eliminated, enabling near-zero crosstalk between the channels.

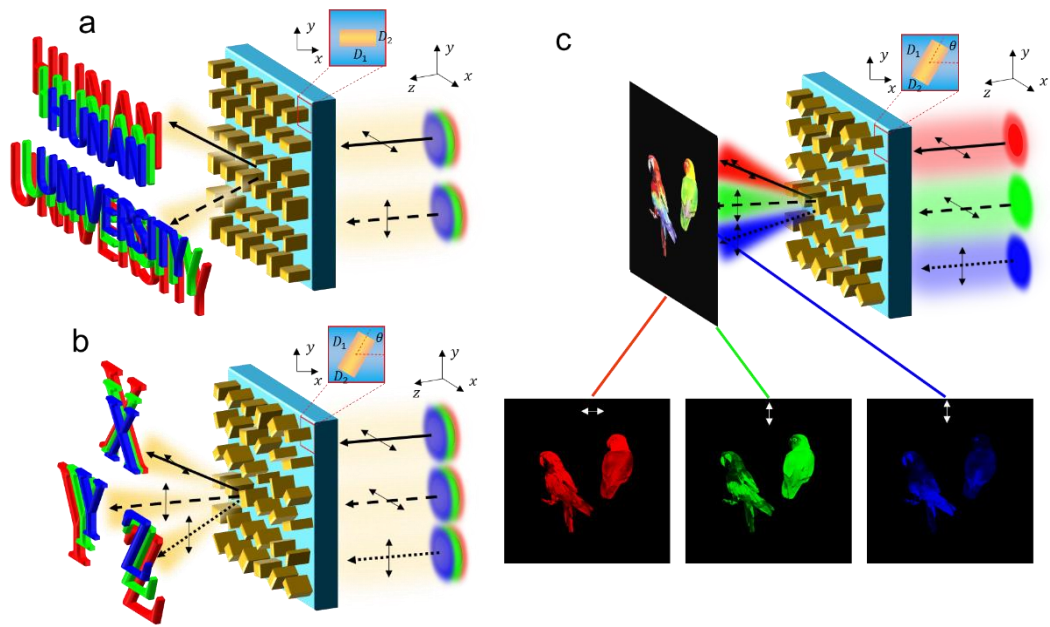

*Figure 1. Schematic of the vectorial full-color dielectric holographic metasurface. (a) Dual-channel polarization-dependent holographic metasurface composed of TiO$_2$ rectangle nanopillars with varied planar dimensions ($D_1$, $D_2$) and zero orientation angle. Two*

*independent holographic images 'HUNAN' and 'UNIVERSITY' can be switched with two orthogonal polarization inputs. (b) Three-channel polarization-dependent holographic metasurface by fully utilizing the structural parameter freedom ($D_1$, $D_2$, $\theta$). Three independent holographic images 'X', 'Y' and 'Z' can be switched under three input/output polarization combinations. The metasurface is broadband in the visible and the field of view (FOV) of holographic images is proportional to the wavelengths. (c) Vectorial full-color holographic metasurface. The grayscale images information of the three primary color wavelengths are coupled into three independent polarization channels and matched by a pre-compensation algorithm to reconstruct a vectorial full-color holographic image. Note that a line with a polarization direction indicates that the input light is polarized or the output light is filtered by specific polarization. The input and output light with the same line type represent as the same light path.*

The metasurfaces are constructed by subwavelength rectangle $TiO_2$ nanopillars. $TiO_2$ is a high refractive index dielectric material in the visible range, which allows easier phase modulation from 0 to $2\pi$. The nanopillars are arranged in squarely repeated units with a period of $p$, which form meta-units. Fig. 2a shows a single meta-unit with three independent tunable structure parameters of $(D_1, D_2, \theta)$ and a fixed height $H$. The rectangle cross-section of the subwavelength nanopillar results in different effective refractive indices along the two axes of the nanopillar. Figure 2b shows the fundamental waveguide modes of the rectangle nanopillar under two orthogonal polarizations. Therefore, each nanopillar can be seen as a linearly birefringent wave plate, which is able to generate distinct phases on orthogonal linear polarizations. If there is the no

amplitude manipulation, the meta-unit can be described using the Jones matrix as

$$J = R(\theta)\begin{bmatrix} e^{i\phi_x} & 0 \\ 0 & e^{i\phi_y} \end{bmatrix} R(-\theta) \quad (1)$$

where $\phi_x$ and $\phi_y$ are the phase shifts on linearly polarized light along its two axes. $R(\theta)$ is the rotation matrix to describe the orientation by an angle of $\theta$ relative to the $x$ axis of the nanopillar. The relation between the electric fields of the input and output light is $\mathbf{E_O} = J\mathbf{E_i}$. Therefore, a single meta-unit can completely control polarization and phase if the freedom of the Jones matrix can be fully utilized by freely choosing the structure parameters.

From Eq.1, we can see that the phase shifts on the $x$- and $y$-polarized light ($\phi_x$, $\phi_y$) are essential to achieve polarization multiplexed function, which should cover $0 - 2\pi$ range. A normally linearly polarized light incidents along the axis of the nanopillar will not change the polarization but generate the phase shift. It means that phase shifts can be expressed as functions of size of the nanopillar $D_1$ and $D_2$. The phase shift $\phi_x$ of the nanopillar with no rotation ($\theta = 0$) under the $x$ polarization were simulated by finite difference time domain (FDTD) method (see Methods) at wavelength of 532 nm, as shown in Fig.2c. The period of the meta-unit was set as 400 nm. The height of the nanopillar was 800 nm to cover about two 0-$2\pi$ ranges so that ensure all possible phase combinations. Similarly, the $\phi_y$ under the $y$ polarization can be simply obtained by transposition of the $x$ polarization results as shown in Fig.2d due to the symmetry. With such databases, any $\phi_x$ and $\phi_y$ combination range in $0 - 2\pi$ can be achieved by properly selecting $D_1$ and $D_2$.

To demonstrate the concept, we considered two schemes of the polarization-dependent multiplexed metasurfaces. First, if meta-units are arranged with no rotation, the Jones matrix can be simplified by $\begin{bmatrix} e^{i\phi_x} & 0 \\ 0 & e^{i\phi_y} \end{bmatrix}$. It means that the output electric field can be obtained by $\begin{bmatrix} e^{i\phi_x} \\ 0 \end{bmatrix}$ and $\begin{bmatrix} 0 \\ e^{i\phi_y} \end{bmatrix}$ under $x$- and $y$-polarized light, respectively. Therefore, two independent phase profile can be encoded into a single metasurface and switched by two orthogonal incident polarizations. Furthermore, to fully utilize the flexibility of the Jones matrix, the three parameters of nanopillars including the rotation angle should be chose freely. Then the output electric field can be expressed by $\frac{\sqrt{2}}{2}\begin{bmatrix} e^{i\phi_1} \\ e^{i\phi_2} \end{bmatrix}$ and $\frac{\sqrt{2}}{2}\begin{bmatrix} e^{i\phi_2} \\ e^{i\phi_3} \end{bmatrix}$ under $x$- and $y$-polarized light, where the phases $\phi_1$, $\phi_2$ and $\phi_3$ are the functions of $\phi_x$, $\phi_y$ and $\theta$. It means that the output electric field contains the components of original and converted orthogonal polarizations. Different circularly polarized light combination results can also be derived. The output light will attach $\phi_1$, $\phi_3$ phase information if the input and output are the same circularly polarization or 45° and 135° polarizations (supplementary section 1). Therefore, seven combinations with the three independent phase profiles can be achieved with different polarization input and output. Then, by solving the inverse problem, the metasurface can transform arbitrary phase and polarization at each pixel range. This gives us a lot of space to design polarization-dependent switchable or dynamic optics. Fig.2e-2h show the top-view and oblique-view scanning electron microscopy (SEM) images of the fabricated two schemes of TiO$_2$ holographic metasurfaces. The devices were fabricated by electron beam lithography (EBL) followed by reactive ion etching (RIE) process.

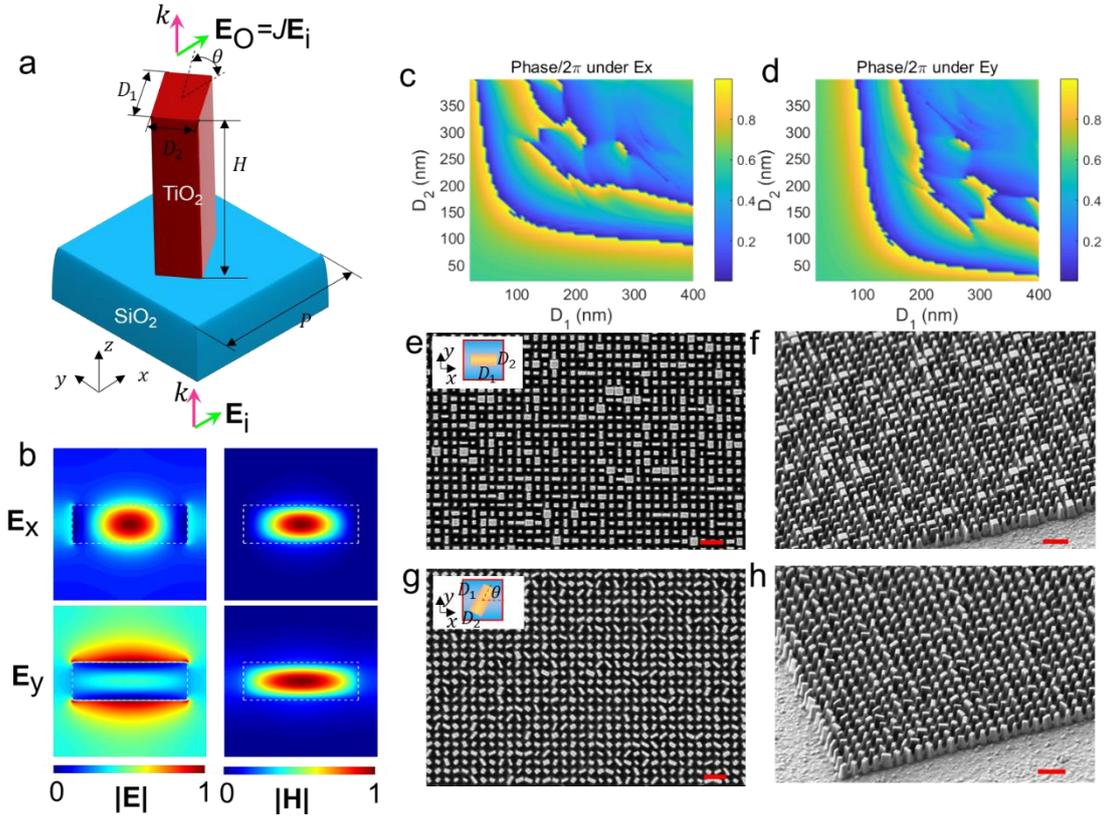

*Figure 2. Vectorial multichannel metasurface design and fabrication.* (a) A single $TiO_2$ meta-unit in a rectangular periodic pixel with three independent tunable structure parameters of $(D_1, D_2, \theta)$ and a fixed height $H$. (b) Distribution of electric and magnetic fields in the meta-unit, showing fundamental waveguide modes of the rectangle nanopillar under two orthogonal polarizations. (c) Simulation results of the phase shift $\phi_x$ of the nanopillar with no rotation ($\theta = 0$) under the $x$ polarization simulated by finite difference time domain (FDTD) method at wavelength of 532 nm. (d) Phase shift $\phi_y$ under the $y$ polarization obtained by transposition of the $x$ polarization results. Scanning electron microscopy (SEM) images of the fabricated (e)(f) dual-channel and (g)(h) three-channel polarization dependent $TiO_2$ holographic metasurfaces.

**Multifunctional Holographic Metasurfaces**

To obtain the phase profiles for hologram images, the Gerchberg–Saxton (GS)

algorithm applied in computer-generated hologram (CGH) was utilized here. A algorithm for solving the inverse problem and matching the structure parameters was also developed (supplementary section 2). The optical experimental setup for capture the hologram and efficiency measurement can be found in supplementary section 4. First, two off-axis images of 'HUNAN' and 'UNIVERSITY' were encoded in a 50 μm ×50 μm metasurface composed of meta-units without rotation (Fig.2e, 2f) to demonstrate the function of the two-channel holographic metasurface, as Fig.3a-3c shown. Two words can be easily switched with $x$- and $y$-polarized light. When the incident light is 45° or circular polarization, two words can be seen simultaneously with same intensity but orthogonal polarizations. Other polarization directions result in two images with different intensity (see the supplementary video 1). The experimental and the simulation results are in good agreement. Here we define the diffraction efficiency as the ratio of the intensity of the single reconstructed image to the power of the transmitted light. The diffraction efficiency of each words was measured to be 55.2% and 49.4% under $x$- and $y$-polarized light in the experiments, respectively. If the incident light was 45° polarized, the diffraction efficiencies were 26.5% and 22.7%, respectively.

In addition, another kind of 50 μm ×50 μm metasurface (Fig.2g and 2h) fully utilizing the parameter freedom by further considering the orientation angle of the meta-units was fabricated to encode three independent information (letters 'X', 'Y' and 'Z'), as shown in Fig. 3d-j. As we mentioned above, the output electric field carries the components of original and converted orthogonal polarizations when this metasurface is shined by a linear polarization. Therefore, different channels can be obtained by filtering the polarization of the output light. First, with different combinations of linear input/output polarizations, three kinds of independent information can be switched, as shown in Fig. 3d-f. Note that letter 'Y' can be obtained by two sets of orthogonal polarizations, but the polarization directions of the output electric field are different. If the output light is not filtered, $x$- and $y$-polarized incident light result in two independent information with orthogonal polarizations (Fig. 3g and 3h). 'X' and 'Z' combination can be obtained if the input and output are the same circularly polarizations or 45° direction (Fig. 3i) because 'Y' carries the polarization direction orthogonal to the incident light. Therefore, when the output electric field is not filtered, the three independent information are displayed at the same time as shown in Fig. 3j. The diffraction efficiency of 'X', 'Y' and 'Z' channel was measured to be 15.8%, 22.0% and 14.5% under 45° polarization, respectively. So far, we have demonstrated that up to three independent information can be encoded into a metasurface composed of noninterleaved meta-units and form up to seven independent combinations. This can be applied for various types of encryption and polarization multi-channel devices. Note that for the figures in the experiments, the zero-order diffraction at the central location was deleted.

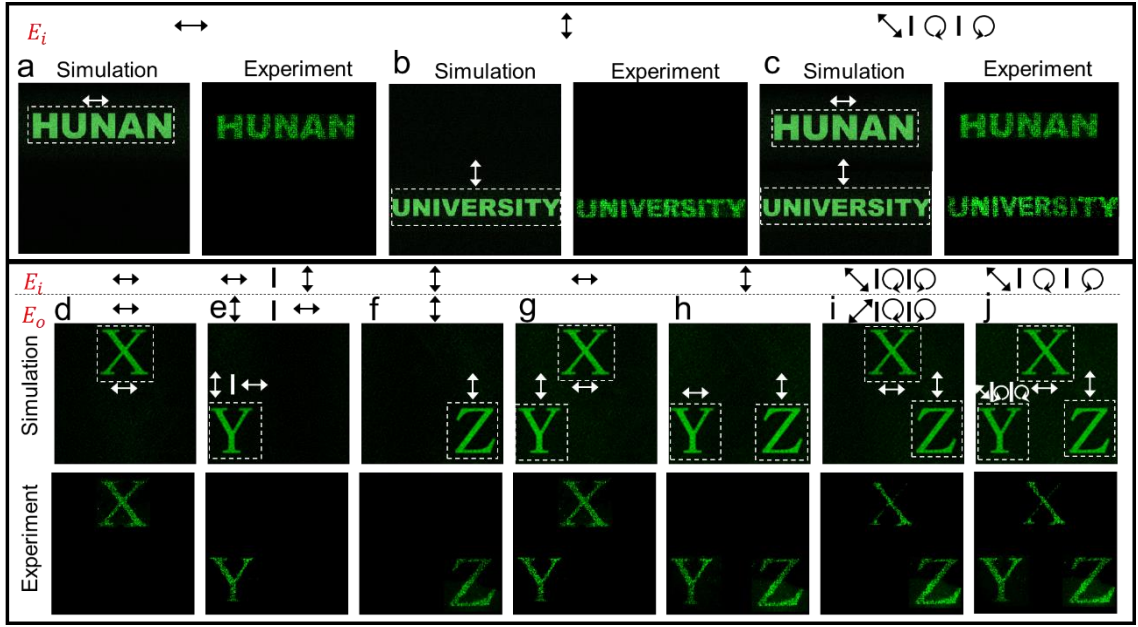

*Figure 3. Optical characterization for vectorial multichannel holographic metasurface. (a)-(c) The reconstructed three independent combination of two vectorial hologram images ('HUNAN' and 'UNIVERSITY') of the vectorial dual-channel holographic metasurface in simulation and experiments under two orthogonal polarizations. (d)-(j) The reconstructed seven independent combinations of three vectorial hologram images ('X', 'Y' and 'Z') of the vectorial three-channel holographic metasurface in simulation and experiments under different input/output polarization combinations.*

**Broadband Property**

In order to achieve full-color holography with noninterleaved metasurface, the metasurface itself must be broadband. We fabricated a 100 μm × 100 μm three-channel metasurface with a designed wavelength of 532 nm and explored the broadband property of the metasurface, as shown in Fig. 4. Three different emoji images were encoded into three polarization channels. Fig. 4a-4b shows that the hologram images can be observed clearly under incident light of R (633 nm), G (532 nm), B (450 nm) wavelengths in the visible range. Note that the size of the hologram image is proportional to the wavelength due to the chromatic aberration.

To understand the broadband property of the metasurface, we first calculated the broadband performance of conversion efficiency of the metasurface, as shown in Fig. 4c. Due to computing resource limit, a 20 μm ×20 μm metasurface encoded with the same information of three emoji images were simulated. The total transmission efficiency $(T_x, T_y)$ and output polarization component efficiency $(T_{xx}, T_{xy}, T_{yx}, T_{yy})$ under $x$ and $y$ polarizations show similar results. Conversion efficiency $T_{xy}, T_{yx}$ shows satisfactory values in the whole visible range, but the component of converted orthogonal polarization decreases when the wavelength is away from the design wavelength. The diffraction efficiency in the experiment of three channels at different wavelengths shows the same trend which are 18.4%, 29.8%, 17.6% (450 nm) and 21.45%, 28.3%, 17.7% (532 nm) and 11.7%, 12.5%, 13.7% (633 nm). This might be because the different phase shifts caused by the changed wavelength, that is, dispersion. To get a better insight of the dispersion effect, we calculated the phase change between $\lambda_i$ and $\lambda_j$ of each meta-unit through waveguide effect as

$$\Delta\phi_{\lambda_i,\lambda_j} = n_i^e \frac{2\pi}{\lambda_i} H - n_j^e \frac{2\pi}{\lambda_j} H \quad (2)$$

where $n_i^e$, $n_j^e$ is the effective refractive index of the fundamental mode of the meta-unit at wavelength of $\lambda_i$ and $\lambda_j$. For different nanopillars in the metasurface, if the phase changes $(\Delta\phi_{\lambda_i,\lambda_j})$ are almost identical, the spatial relative phase distribution remains unchanged, leading to the same reconstructed holographic information. From Eq. 2, we can see that structural dispersion (that is, different structural parameters cause different effective refractive indices) might result in phase mismatch. We numerically calculated the effective modal indices of the waveguide modes using eigenmode analysis (See methods). Note that we ignored the power coupling into higher-order waveguide modes and estimated the effective indices based only on the fundamental waveguide modes since high-order modes have little effect on the phase response [33]. Fig.4d and 4e shows the phase changes compared to the designed wavelength of 532 nm and the statistical distribution of the number of meta-units. We can see that most of the meta-units locate in the region where the phase changes are almost the same. Note that the area with close phase difference of $\Delta\phi_{450,532}$ is bigger than that of $\Delta\phi_{532,633}$, explaining why the conversion efficiency is lower at 633 nm in Fig.4c.

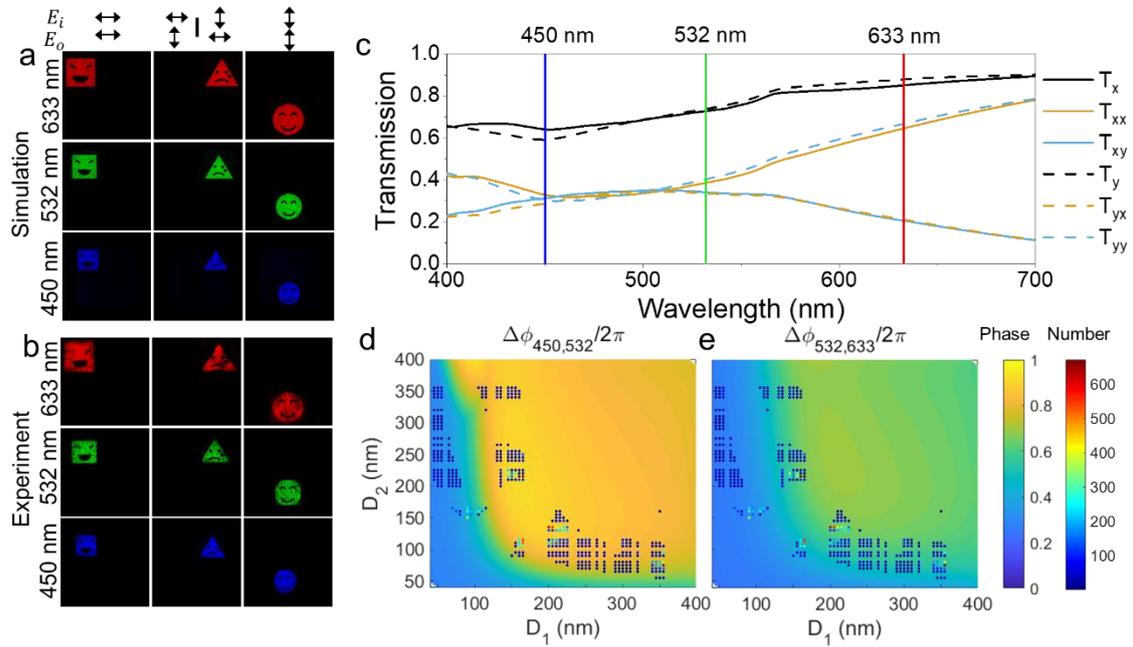

*Figure 4. Broadband Property of the vectorial multichannel metasurface.* *(a)-(b) Three independent reconstructed hologram images (three emoji) under 633 nm, 532 nm and 450 nm wavelengths of three-channel holographic metasurface in simulation and experiments. (c)*

*Transmission and conversion efficiency of the metasurface. (d, e) The phase changes compared to the design wavelength of 532 nm and the statistical distribution of the number of meta-units.*

**Vectorial Trichromatic Hologram**

It has been proved above that the proposed metasurface can support three kinds of independent polarization-dependent information and is broadband, which can be combined with the principle of three primary colors to realize vectorial full color holography. Fig. 5a shows the generation process of the vectorial color holographic metasurface. First, the target color image (e.g. a painting of 'Two parrots') was divided into three trichromatic components corresponding to wavelengths of 633 nm (R), 532 nm (G) and 450 nm (B). Due to the chromatic aberration, the three components were scaled inversely proportional to the wavelength to match the size in the projected screen. In addition, the components were pre-compensated for each wavelength to eliminate the distortion due to the large field of view (FOV) (Supplementary section 3). After that, the GS algorithm was applied to generate the phase profiles of the pre-compensated components, which then were coupled into three independent polarization channels. Based on the phase distribution, the structural parameters and orientation angles corresponding to the structure with the closest phase response were found to form the layout of the metasurface. A 400 μm × 400 μm metasurface encoded with the information of the color painting was then fabricated and characterized (the setup can be found in the Fig. S4b). The experimental results in Fig. 5g-5j show a satisfied agreement with the simulation results. Since the input/output polarization combination of each channel is independent, it can be seen that there is almost no crosstalk between each channel. We can see that for the case where the input and output polarizations are orthogonal (i.e. G channel), the signal-to-noise ratio (SNR) is better than that with the same input/ output polarizations (i.e. R and B channels) due to the background light. Fig. 5k-5f shows the partially enlarged view of the hologram image in the simulation and experiments. It can be seen that the result in experiments restores the color well, further demonstrating that the polarization multi-channel hologram can extremely suppress crosstalk among the channels.

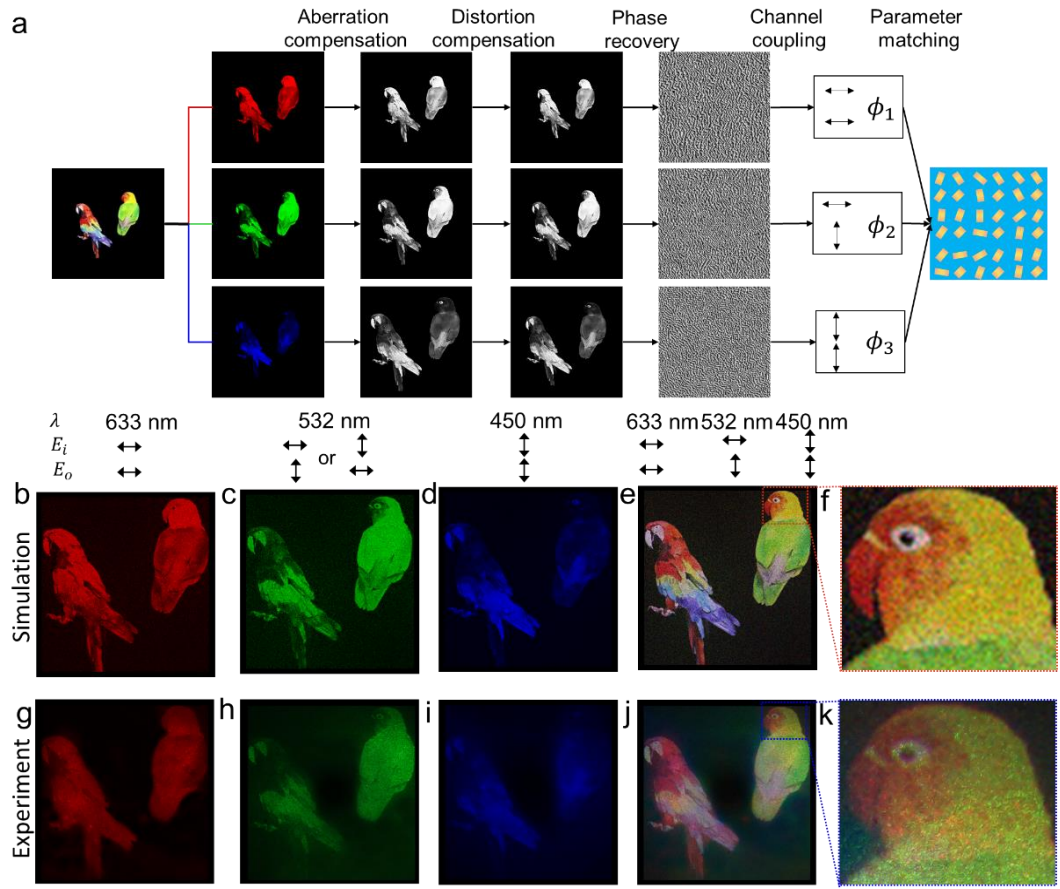

*Figure 5. Vectorial Trichromatic Meta-Holography.* *(a) Design process for vectorial color holographic metasurface. (b)-(e) Simulation and experimental results of the three components and channel mixing of the vectorial color meta-holography. (f)-(k) Enlarged view of a partial of the full-color meta-hologram.*

**Discussion**

In fact, the decoupling of wavelength three-channel and polarization three-channel enables the device to encode 9 kinds of independent information with extra spatial freedom to reconstruct $2^9-1=511$ polarization- and wavelength- dependent information combinations or up to three color hologram images which can be widely applied for dynamic display, multi-task devices, optical encryption and data storage. Therefore, the regulation of arbitrary polarization in a single pixel is the irreplaceable advantage of the metasurface. Based on this, recently, the polarization imaging [34], polarization detection [35] and arbitrary spin-to-orbital angular momentum convertor [36] have been proposed. It will continue to expand into various fields such as optical communication, display, and imaging.

In summary, we utilized the full degree of freedom of a single meta-unit to achieve a high efficiency vectorial metasurface encoded with up to three kinds of independent information and seven independent information combinations in the visible light range. Combining three polarization channels and three primary color channels ingeniously, a full-color hologram with near-zero crosstalk was realized for the first time. Our proposed polarization-dependent multi-channel vectorial holography can be used for data encryption and anti-counterfeiting. In addition, compared to the color meta-holography based on the interleaved meta-units, the proposed high-quality color holography has made a big step towards the practical application, which may be widely used in the field of near-eye display, holographic display, naked-eye three-dimensional display.

## Methods

**Numerical simulation.** The phase responses and transmission efficiency of the metasurface were simulated by the finite-difference time-domain (FDTD) method (Lumerical FDTD Solutions). A 2D parameter sweep of the a single meta-unit with length and width from 40 nm – 400 nm and fixed height 800 nm under $x$ polarization was carried out to obtain the phase map of the different structural parameters in Fig. 2d. The phase response under $y$ polarization was obtained by transposition of the $x$ polarization results. The refractive index of the $TiO_2$ is the measurement result by ellipsometer. The waveguide mode of the meta-unit was carried out by the eigenmode analysis (Lumerical MODE Solutions). The effective refractive index is obtained from the fundamental modes while the power coupling into higher-order waveguide modes was ignored.

**Optical characterization.** The optical setup for full color hologram characterization is shown in Fig. S4. Three laser diodes emitting at 450 nm, 532 nm and 633 nm were utilized as R, G and B channels. A combination of linear polarizer and quarter-wave plate was placed behind each laser to produce arbitrary polarizations. Then two dichroic lenses with cutoff wavelength of 567 nm and 490 nm were used to combine the three lasers in one light path. The metasurface was placed on the focal plane of an objective lens (×50/NA = 0.75) to ensure that the Fourier plane is located in the back focal plane and the numerical aperture (NA) was chosen to collect all the diffraction light from the metasurface. Another dichroic lens at 490 nm was placed right behind the objective lens to separate B channel and RG channel, which then were both filtered by another combination of linear polarizer and quarter-wave plate with desired polarizations. Two mirrors and dichroic lens were used to combine the three channel again. A lens was used for capturing the Fourier plane on a CCD camera. For monochrome hologram, one of three lasers was turned on.

**Acknowledgments**

We acknowledge the financial support by the National Natural Science Foundation of China under contract no. 51722503, 51621004 and 11574078. C.W.Q. acknowledges the financial support from the National Research Foundation, Prime Minister's Office, Singapore under its Competitive Research Program (CRP award NRF CRP15-2015-03).

**Author contributions:**

Y.H., H.D. and L.L. proposed the idea. Y.H., L.L. conceived and carried out the design and simulation. Y.H., M.M. X.L., Y. C., H.W. and Y.L. fabricated the samples. Y.H. L.J. and X.L. conceived and performed the measurements. Y.H., H.D., L.L., L.J. and C.-W. Q. discussed the results and co-wrote the manuscript. H.D. supervised the overall project. All the authors discussed the results and commented on the manuscript.